%% file: main.tex
\PassOptionsToPackage{hyphens}{url}
\documentclass[camera]{jpaper}

\usepackage{amsmath,amssymb}
\usepackage[binary-units=true]{siunitx}
\usepackage{graphicx}
\usepackage{url}
\usepackage{fancyhdr}
\usepackage{xspace}
\usepackage{graphicx}
\usepackage{url}
\usepackage{algorithm}
\usepackage{algorithmic}
\usepackage{array}
\usepackage{wrapfig}
\usepackage{tikz}
\usepackage{amsmath}
\usepackage{titlesec}
\usepackage[nocompress]{cite}
\usepackage{setspace}
\usepackage{footmisc}
\usepackage{listings}
\usepackage{dblfloatfix} 
\usepackage{courier}
\usepackage{indentfirst}
\usepackage{fixltx2e}
\usepackage[nolessnomore, italic]{mathastext}
\usepackage[T1]{fontenc}
\usepackage[usenames,dvipsnames,svgnames,table]{xcolor}
\usepackage[normalem]{ulem}
\usepackage{enumitem}
\usepackage{authblk}
\usepackage[us,12hr]{datetime}
\usepackage[keeplastbox]{flushend}
\usepackage[hidelinks]{hyperref}

\makeatletter
\let\MYcaption\@makecaption
\makeatother

\makeatletter
\let\@makecaption\MYcaption
\makeatother

\usepackage{subfig}

\newcommand*\circled[1]{\tikz[baseline=(char.base)]{
            \noindent
            \node[shape=circle,draw,fill=white,text=black,inner sep=0.5pt] (char) {#1};}}

\newcommand{\ignore}[1]{}

\newcommand{\note}[1]{}
\newcommand{\squishlist}{
   \begin{list}{$\bullet$}
    { \setlength{\itemsep}{0pt}    \setlength{\parsep}{0pt}
      \setlength{\topsep}{0pt}     \setlength{\partopsep}{0pt}
      \setlength{\leftmargin}{2em} \setlength{\labelwidth}{1.5em}
      \setlength{\labelsep}{0.5em} } }

\newcommand{\squishend}{
    \end{list} }

\newcommand{\trcd}{\texttt{{tRCD}}\xspace}
\newcommand{\tras}{\texttt{{tRAS}}\xspace}
\newcommand{\trp}{\texttt{{tRP}}\xspace}
\newcommand{\twr}{\texttt{{tWR}}\xspace}

\newcommand{\trefi}{\texttt{{tREFI}}\xspace}

\newcommand{\cmdact}{\texttt{{ACTIVATE}}\xspace}
\newcommand{\cmdread}{\texttt{{READ}}\xspace}
\newcommand{\cmdwrite}{\texttt{{WRITE}}\xspace}
\newcommand{\cmdprech}{\texttt{{PRECHARGE}}\xspace}

\newif\ifcameraready
\camerareadytrue

\newcommand{\versionnum}[0]{5}

\ifcameraready
\else
\fi

\newcommand{\ch}[1]{{#1}\xspace}
\newcommand{\chs}[1]{{#1}\xspace} 
\ifcameraready
  \newcommand{\todo}[1][]{}
  \newcommand{\chI}[0]{}
\else
  \newcommand{\todo}[1][]{\textbf{\fcolorbox{black}{red}{\color{white}{TODO}}} \underline{$\overline{\hbox{\emph{#1}}}$}}
  \newcommand{\chI}[1]{{\color{BrickRed} #1}}
\fi

\begin{document}

\title{SoftMC: Practical DRAM Characterization \\ Using an FPGA-Based Infrastructure}

\author{
{Hasan Hassan$^{1,2,3}$}%
\qquad%
{Nandita Vijaykumar$^{2}$}%
\qquad%
{Samira Khan$^{4,2}$}%
\vspace{2pt}\\%
{Saugata Ghose$^{2}$}%
\qquad%
{Kevin Chang$^{5,2}$}%
\qquad%
{Gennady Pekhimenko$^{6,2}$}%
\vspace{2pt}\\%
{Donghyuk Lee$^{7,2}$}%
\qquad%
{Oguz Ergin$^{3}$}%
\qquad%
{Onur Mutlu$^{1,2}$}%
}
\affil{
{\it%
$^1$ETH Z{\"u}rich\qquad%
$^2$Carnegie Mellon University\qquad%
$^3$TOBB University of Economics \& Technology}%
\vspace{2pt}\\%
{\it%
$^4$University of Virginia\qquad
$^5$Facebook\qquad
$^6$Microsoft Research\qquad
$^7$NVIDIA Research}%
}

\maketitle

\input{sections/abstract}
\input{sections/summary}
\input{sections/use_cases_1}

\input{sections/use_cases_2}
\input{sections/related_work}

\input{sections/significance}

\input{sections/conclusion}

\input{sections/acknowledgements}

{
\bibliographystyle{IEEEtranS}
\bibliography{refs}
}

\end{document}

%% file: sections/abstract.tex
\begin{abstract}

\chs{This paper summarizes the SoftMC DRAM characterization infrastructure, which was published in
HPCA 2017~\cite{hassan2017softmc}, and examines the work's significance and
future potential.} 
DRAM is the primary technology used for main memory in
modern systems. Unfortunately, as DRAM scales down to smaller technology nodes,
it faces key challenges in both data integrity and latency, which strongly
\ch{affect} overall system reliability and performance.  To develop reliable
and high-performance DRAM-based main memory in future systems, it is critical
to characterize, understand, and analyze various aspects (e.g., reliability,
latency) of \ch{modern} DRAM chips. To enable this, there is a strong need for
a publicly-available DRAM testing infrastructure that can flexibly and
efficiently test DRAM chips in a manner accessible to both software and
hardware developers.
 
    This \ch{work} develops the first such infrastructure, {\em SoftMC (Soft
Memory Controller)}, an FPGA-based testing platform that can control and test
memory modules designed for the commonly-used DDR (Double Data Rate) interface.
SoftMC has two key properties: {\em (i)} it provides {\em flexibility} to
thoroughly control memory behavior or to implement a wide range of mechanisms
using DDR commands; and {\em (ii)} it is {\em easy to use} as it provides a
simple and intuitive high-level programming interface for users, completely
hiding the low-level details of the FPGA.

We demonstrate the capability, flexibility, and programming ease of SoftMC
    with two example use cases. First, we implement a test that
    characterizes the retention time of DRAM cells. Experimental results we
    obtain using SoftMC are consistent with the findings of prior studies
    on retention time in modern DRAM, which serves as a validation of our
    infrastructure. Second, we validate two recently-proposed mechanisms,
    which rely on accessing recently-refreshed or recently-accessed DRAM
    cells faster than other DRAM cells.
    Using our infrastructure, we show that the expected latency
    reduction effect of these mechanisms is not observable in existing DRAM
    chips,
%
 which demonstrates the usefulness of SoftMC in testing new ideas on
    existing memory modules.  
    
        \chs{Various versions of the SoftMC platform have enabled many of our other
    DRAM characterization studies~\cite{chang2016understanding, khan2016parbor, khan2016case,
kim2014flipping, khan-sigmetrics2014, liu2013experimental,
lee-hpca2015, lee2017design, chang2017understanding, qureshi-dsn2015}.}
        We discuss several other use cases of SoftMC,
    including the ability to characterize emerging non-volatile memory
    modules that obey the DDR standard. 
    We hope that our open-source
    release of SoftMC fills a gap in the space of publicly-available
    experimental memory testing infrastructures and inspires new studies,
    ideas, and methodologies in memory system design.

\end{abstract}

%% file: sections/summary.tex


\section{Understanding DRAM Characteristics}

DRAM (Dynamic Random Access Memory) is the predominant technology used
to build main memory systems of modern computers. The continued
scaling of DRAM process technology has enabled tremendous growth in
DRAM density in the last few decades, leading to higher capacity
main memories. Unfortunately, as the process technology node scales down to
the sub-\SI{20}{\nano\meter} feature size range, DRAM technology faces key challenges
that critically impact its reliability and
performance\ch{~\cite{mutluresearch, mutlu2013memory, mutlu2017rowhammer}}.

The fundamental challenge with scaling DRAM cells into smaller technology
nodes arises from the way DRAM stores data in cells. A DRAM cell consists
of a transistor and a capacitor. Data is stored as charge in the capacitor.
A DRAM cell cannot retain its data permanently as this capacitor leaks its
charge gradually over time.  To maintain correct data in DRAM, each cell is
periodically refreshed to replenish the charge in the
capacitor\ch{~\cite{liu2012raidr}}. At
smaller technology nodes, it is becoming increasingly difficult to store
and retain enough charge in a cell, causing various reliability and
performance issues\ch{~\cite{liu2012raidr, liu2013experimental,
chang2014improving, khan2017detecting}}.  Ensuring
reliable operation of the DRAM cells is a key challenge in future
technology nodes\ch{~\cite{mutlu2013memory, mandelman2002challenges,
kim2005technology, mueller2005challenges, liu2012raidr, kang-memcon2014,
liu2013experimental, patel2017reach, mutlu2017rowhammer,
khan-sigmetrics2014}}.

The fundamental problem of retaining data with less charge in smaller cells
directly impacts the reliability and performance of DRAM cells.  First,
smaller cells placed in close proximity make cells more susceptible to
various types of interference. This potentially disrupts DRAM operation by
flipping bits in DRAM, resulting in major reliability
issues\ch{~\cite{kim2014flipping, nakagome1988impact,
redeker2002investigation, schroeder2009dram, sridharan2013feng,
meza-dsn2015, sridharan2015memory}}, which can lead to system
failure~\cite{schroeder2009dram,meza-dsn2015} or security
breaches\ch{~\cite{kim2014flipping, seaborn2015exploiting,
seaborn2015exploiting2, rowhammerjava2015, van2016drammer, xiao2016one,
razavi2016flip, bosman2016dedup}}. Second, it takes longer time to access a
cell with less charge~\cite{hassan2016chargecache, lee-hpca2015}, and write
latency increases as the access transistor size
reduces~\cite{kang-memcon2014}. Thus, smaller cells directly impact DRAM
latency, as DRAM access latency is determined by the worst-case (i.e., {\em
slowest}) cell in any \ch{acceptable}
chip\ch{~\cite{chandrasekar2014exploiting, lee-hpca2015,
chang2017understanding}}. DRAM access latency has not \ch{significantly}
improved with technology scaling in the past \ch{two decades
~\cite{lee-hpca2013, jung-2005, borkar-cacm2011, mutlu2013memory,
chang2016understanding, lee2016thesis, chang2017thesis}}, and, in fact,
some latencies are expected to increase~\cite{kang-memcon2014}, making
memory latency an increasingly critical system performance bottleneck.

As such, there is a significant need for new mechanisms that improve the
reliability and performance of DRAM-based main memory systems. In order to
design, evaluate, and validate many such mechanisms, it is important to
accurately characterize, analyze, and understand DRAM (cell) behavior in
terms of reliability and latency. For such an understanding to be accurate,
it is critical that the characterization and analysis be based on the
\emph{experimental} studies of {\em real DRAM chips}, since a large number
of factors (e.g., various types of cell-to-cell
interference~\cite{nakagome1988impact, redeker2002investigation,
kim2014flipping}, inter- and intra-die process
variation\ch{~\cite{chang2016understanding, nassif2000delay,
chandrasekar2014exploiting, lee-hpca2015, chang2017understanding,
lee2017design, patel2017reach, kim2018dram}}, random
effects\ch{~\cite{liu2013experimental, yaney1987meta, srinivasan1994accurate,
hazucha2000impact, restle1992dram, khan-sigmetrics2014, qureshi-dsn2015}},
operating conditions\ch{~\cite{liu2013experimental,
lee2010mechanism, lee-hpca2015, chang2017understanding, patel2017reach,
kim2018dram}}, internal organization~\cite{liu2013experimental,
hidaka1989twisted, khan2016parbor}, stored data
patterns~\cite{khan2016case, khan2016parbor, liu2013experimental})
concurrently impact the reliability and latency of cells. Many of these
phenomena and their interactions cannot be properly modeled (e.g., in
simulation or using analytical methods) without rigorous experimental
characterization and analysis of real DRAM chips.  The need for such
experimental characterization and analysis, with the goal of building the
understanding necessary to improve the reliability and performance of
future DRAM-based main memories at various levels (both software and
hardware), motivates the need for a publicly-available DRAM testing
infrastructure that can enable system users and designers to characterize
real DRAM chips.

%
%

\section{Experimental DRAM Characterization}

Two key features are desirable from an experimental memory testing
infrastructure. First, the infrastructure should be {\em flexible} enough
to test any DRAM operation (supported by the commonly-used DRAM interfaces,
e.g., the standard Double Data Rate, or DDR, interface) to characterize
cell behavior or evaluate the impact of a mechanism (e.g., adopting
different refresh rates for different cells\ch{~\cite{liu2012raidr,
venkatesan2006retention, khan-sigmetrics2014, qureshi-dsn2015,
khan2016case, khan2017detecting, patel2017reach}}) on real DRAM chips.
Second, the infrastructure should be {\em easy to use}, such that it is
possible for both software and hardware developers to implement new tests
or mechanisms without spending significant time and effort. For example, a
testing infrastructure that requires circuit-level implementation, detailed
knowledge of the physical implementation of DRAM data transfer protocols
over the memory channel, or low-level FPGA-programming to modify the
infrastructure would severely limit the usability of such a platform to a
limited number of experts.

Our HPCA 2017 paper~\cite{hassan2017softmc} designs, prototypes, and
demonstrates the basic capabilities of such a flexible and easy-to-use
experimental DRAM testing infrastructure, called {\em SoftMC (Soft Memory
Controller)}. SoftMC is an open-source FPGA-based DRAM testing
infrastructure, consisting of a programmable memory controller that can
control and test memory modules designed for the commonly-used DDR (Double
Data Rate) interface. To this end, SoftMC implements {\em all} low-level
DRAM operations (i.e., DDR commands) available in a typical memory
controller (e.g., opening a row in a bank, reading a specific column
address, performing a refresh operation, enforcing various timing
constraints between commands).  Using these low-level operations, SoftMC
can test and characterize any (existing or new) DRAM mechanism that uses
the existing DDR interface. SoftMC provides a simple and intuitive
high-level programming interface that completely hides the low-level
details of the FPGA from users. Users implement their test routines or
mechanisms in a high-level language that automatically gets translated into
the low-level SoftMC memory controller operations in the FPGA.

\section{Overview \ch{of SoftMC}}

A publicly-available DRAM testing infrastructure 
should have two
key features to ensure widespread adoption among architects and designers:
{\em (i)}~flexibility and {\em (ii)}~ease of use.

\textbf{Flexibility.} 
A DRAM \ch{chip} is typically accessed by issuing a set of DRAM commands in a
particular sequence with a strict delay between the commands (specified by
the timing parameters in the datasheet of the DRAM \ch{chip/module}). A DRAM testing
infrastructure should implement all low-level DRAM operations with tunable
timing parameters without any restriction on the ordering of DRAM commands.
Such a design enables flexibility at two levels. First, it enables
comprehensive testing of \emph{any} DRAM operation with the ability to
customize the length of each timing constraint. For example, we can
implement a retention test with different refresh intervals to characterize
the distribution of retention time in modern DRAM chips\ch{~(as done
in~\cite{liu2012raidr, patel2017reach, khan-sigmetrics2014})}. Such a
characterization can enable new mechanisms to reduce the number of refresh
operations in DRAM, leading to performance and power efficiency
improvements. Second, it enables testing of DRAM chips with high-level test
programs, which can consist of \emph{any combination of DRAM operations and
timings}. Such flexibility is extremely powerful to test the impact of
existing or new DRAM mechanisms in real DRAM chips.

\textbf{Ease of Use.} A DRAM testing infrastructure should provide a simple
and intuitive programming interface that minimizes programming effort and
time. An interface that hides the details of the underlying implementation
is accessible to a wide range of users. With such a high-level abstraction,
even users that lack hardware design experience should be able to develop
DRAM tests.

\begin{figure}[!ht] \centering
    \includegraphics[width=0.53\linewidth]{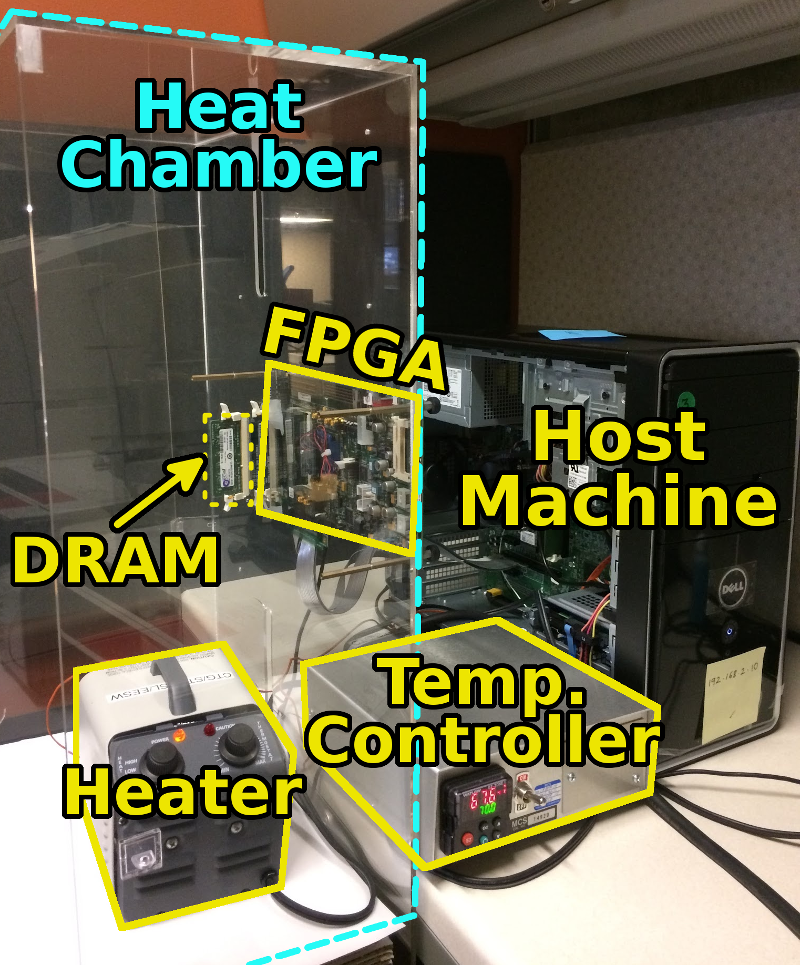}%
    \caption{Our SoftMC infrastructure. \ch{Reproduced
    from~\cite{hassan2017softmc}}\chs{.}}
    \label{fig:softmc_overview}
\end{figure}

Figure~\ref{fig:softmc_overview} shows our temperature-controller
setup for testing DRAM modules. The components of SoftMC operate on the
\emph{host machine} and the \emph{FPGA}. On the host machine, the
\emph{SoftMC API} provides a high-level software interface (in C++) for
developing a test program that generates DRAM commands and sends them to
the FPGA. On the FPGA, \emph{SoftMC hardware} is responsible for handling
the commands sent by the host machine. The SoftMC hardware issues the DRAM
commands in order and with the timing parameters as defined in the test
program developed using the SoftMC API. SoftMC also implements a PCIe
driver for high-speed communication between the host machine and the FPGA.
The user only needs to focus on defining a routine for testing the DRAM.

\ch{
    A detailed description of the interface, design, and \chs{operation} of
SoftMC can be found in our HPCA 2017 paper~\cite{hassan2017softmc}. The
source code for SoftMC can be freely downloaded from~\cite{safarigithub}.
}

%% file: sections/use_cases_1.tex
\section{Example Use Cases}
\label{sec:use_cases}
Using our SoftMC prototype, we perform two case studies on
randomly-selected real DRAM chips from three \ch{major} manufacturers. First, we
discuss how a simple retention test can be implemented using SoftMC, and
present the experimental results of that test (Section~\ref{sec:usecase1}).
Second, we demonstrate how SoftMC can be leveraged to test the expected
effect of two recently-proposed mechanisms\ch{~\cite{hassan2016chargecache,
shin2014nuat}} that aim to reduce DRAM access
latency (Section~\ref{sec:nuat_test}). Both use cases demonstrate the
flexibility and ease of use of SoftMC.


\subsection{Retention Time Distribution Study}
\label{sec:usecase1}

This test aims to characterize data retention time in different DRAM
modules. The retention time of a cell can be determined by testing the cell
with different refresh intervals. The cell fails at a refresh interval that
is greater than its retention time. In this test, we gradually increase
the refresh interval from the default \SI{64}{\milli\second} and count the
number of bytes that have an incorrect value at each refresh interval.

\subsubsection{Evaluating Retention Time with SoftMC}
\label{sec:ret_test_method}

We perform a simple test to measure the retention time of the cells in a
DRAM chip. Our test consists of three steps: {\em(i)} We write a reference
data pattern (e.g. all zeros, or all ones) to an entire row. 
{\em (ii)} We wait for the specified refresh interval, so that the row is
idle for that time and all cells gradually leak charge. {\em (iii)} We read
data back from the same row and compare it against the reference pattern
that we wrote in the first step.  Any mismatch indicates that the cell
could not hold its data for that duration\chs{,} \ch{resulting} in a bit flip. We
count the number of bytes that have bit flips for each test. 

We repeat this procedure for all rows in the DRAM module. The read and
write operations in the test are issued with the standard timing
parameters, to make sure that the only \ch{timing change} that affects the
\ch{reliability of the} cells is the \ch{change in the} refresh interval.



\textbf{Writing Data to DRAM.} 
In Program~\ref{code:write_row}, we present the implementation of the first
part of our retention time test, where we write data to a row, using the
SoftMC API. First, to activate the row, we insert the instruction generated
by the \textit{genACT()} function to an instance of the
\emph{InstructionSequence} (Lines~1-2). This function is followed by a
\textit{genWAIT()} function (Line~3) that ensures that the activation
completes with the standard timing parameter \trcd. Second, we issue write
instructions to write the data pattern in each column of the row. This is
implemented in a loop, where, in each iteration, we call \textit{genWR()}
(Line~5), followed by a call to \textit{genWAIT()} function (Line~6) that
ensures proper delay between two \cmdwrite operations. After writing to all
columns of the row, we insert another delay (Line~8) to account for the
\emph{write recovery} time \twr. Third, once we have written to all
columns, we close the row by precharging it. This is done by the
\textit{genPRE()} function (Line~9), followed by a \textit{genWAIT()}
function with standard \trp timing.\ch{\footnote{For details on DRAM timing
parameters and internal DRAM operation, we refer the reader to our prior
works~\cite{kim-isca2012, lee-hpca2013, lee-hpca2015, seshadri2013,
chang2016understanding, chang2017understanding, chang2014improving,
chang2016, seshadri2017ambit, lee2017design, kim2010atlas, kim2010thread,
hassan2016chargecache, liu2012raidr, liu2013experimental, hassan2017softmc,
kim2015ramulator, lee-pact2015, lee-taco2016, kim2014flipping,
patel2017reach, kim2018dram}.}} Finally, we call the \textit{genEND()}
function to indicate the end of the instruction sequence, and send the test
program to the FPGA by calling the \textit{execute()} function.


\begin{lstlisting}[%float, %linewidth=.9\linewidth,
 language=C, captionpos=b, caption=Writing data
to a row using the SoftMC API.   \ch{Reproduced
    from~\cite{hassan2017softmc}}\chs{.},
label=code:write_row,
numbers=left, xleftmargin=1.75\parindent, xrightmargin=-1.75\parindent, framexrightmargin=-2.1\parindent]
InstructionSequence iseq;
iseq.insert(genACT(bank, row));
iseq.insert(genWAIT(tRCD));
for(int col = 0; col < COLUMNS; col++){
    iseq.insert(genWR(bank, col, data));
    iseq.insert(genWAIT(tBL));
}
iseq.insert(genWAIT(tCL + tWR));
iseq.insert(genPRE(bank));
iseq.insert(genWAIT(tRP));
iseq.insert(genEND());
iseq.execute(fpga));
\end{lstlisting}


\textbf{Employing a Specific Refresh Interval.} 
Using SoftMC, we can implement the target refresh interval in two ways. 
We can use the auto-refresh support provided by the SoftMC
hardware, by setting the \trefi parameter to our target value,
and letting the FPGA take care of the refresh operations.
Alternatively, we can disable auto-refresh, 
and manually control the refresh operations from the software. In this case,
the user is responsible for issuing refresh operations at the right time.
In this retention test, we disable auto-refresh
and use a software clock to determine when we should read
back data from the row (i.e., refresh the row).

\textbf{Reading Data from DRAM.} Reading data back from the DRAM requires
steps similar to DRAM writes (presented in Program~\ref{code:write_row}).
The only difference is that, instead of issuing a \cmdwrite command, we
need to issue a \cmdread command and enforce read-related timing
parameters. In the SoftMC API, this is done by calling the \textit{genRD()}
function in place of the \textit{genWR()} function, and specifying the
appropriate read-related timing parameters. After the read operation is
done, the FPGA sends back the data read from the DRAM module, and the user
can access that data using the \textit{fpga\_recv()} function provided by
the driver.


\ch{Note that the complete code to implement our full retention test (i.e.,
writing a data pattern to \chs{a} DRAM module, waiting for the target retention
time, reading the data back from the DRAM module, and checking the data for
errors) in SoftMC takes \chI{\emph{only} approximately} 200 lines of C code, in the form shown
in Program~\ref{code:write_row}\chs{.}} Based on the intuitive code implementation
of the retention test, we conclude that it requires minimal effort to write
test programs using the SoftMC API. Our full test is provided in our
open-source release \ch{of SoftMC}~\cite{safarigithub}.

\subsubsection{Results}
\begin{figure}[!hb] \centering
\includegraphics[width=.98\linewidth]{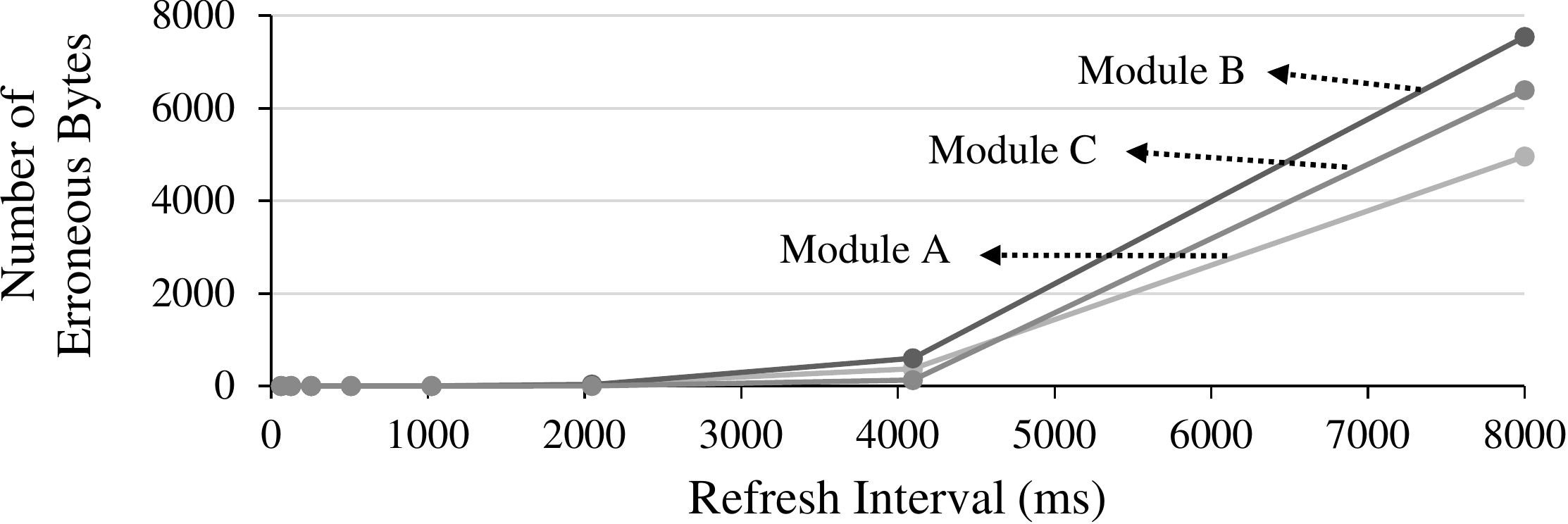}
\caption{Number of erroneous bytes observed in retention time tests.
    Reproduced from~\cite{hassan2017softmc}\ch{.}}
\label{fig:ret_time_test}
\end{figure}

We perform the retention time test at room temperature, using 24 \ch{DRAM} chips from
three major manufacturers. We vary the refresh interval from
\SI{64}{\milli\second} to \SI{8192}{\milli\second}\ch{,} exponentially.
Figure~\ref{fig:ret_time_test} shows the results for the test, where the
x-axis shows the refresh interval in milliseconds, and the y-axis shows the
number of erroneous bytes found in each interval. We make two major
observations. 

{\em (i)} We do not observe any retention failures until we test with a
refresh interval of \SI{1}{\second}. This shows that there is a large
safety margin for the refresh interval in modern DRAM chips, which is
conservatively set to \SI{64}{\milli\second} by the DDR
standard.\footnote{DRAM manufacturers perform retention tests that are
similar to ours (but with proprietary in-house infrastructures that are not
disclosed). Their results are similar to
ours\ch{~\cite{liu2013experimental, lee-hpca2015, khan-sigmetrics2014,
chang2016understanding, patel2017reach, kim2009new,
hamamoto1998retention}}, showing significant margin for the refresh
interval.  This margin is added to ensure reliable DRAM operation for the
worst-case operating conditions (i.e., worst case temperature \ch{and
voltage levels}) and for worst-case cells, as has been shown by prior
works\ch{~\cite{lee-hpca2015, liu2013experimental, khan-sigmetrics2014,
chang2016understanding, patel2017reach, kim2009new,
hamamoto1998retention}}.}


{\em(ii)} We observe that the number of failures increases exponentially 
with the increase in refresh interval.

\ch{Other} experimental studies on retention time of DRAM cells have
reported similar observations as ours\ch{~\cite{khan-sigmetrics2014,
liu2013experimental, lee-hpca2015, hou2013fpga, hamamoto1998retention,
patel2017reach, kim2009new, lee-hpca2015}}. We conclude that SoftMC can
easily reproduce experimental \ch{DRAM} results, validating the
correctness of our testing infrastructure and showing its flexibility and
ease of use.

%% file: sections/use_cases_2.tex
\subsection{Evaluating the Expected Effect of Two Recently-Proposed
Mechanisms in Existing DRAM Chips}
\label{sec:nuat_test}

Two recently-proposed mechanisms, ChargeCache~\cite{hassan2016chargecache}
and NUAT~\cite{shin2014nuat}, provide low-latency access to highly-charged
DRAM cells. They both are based on the key idea that a
highly-charged cell can be accessed faster than a cell with less
charge~\cite{lee-hpca2015}. ChargeCache observes that cells
belonging to \emph{recently-accessed} DRAM rows are in a highly-charged
state and that such rows are likely to be accessed again
in the near future. ChargeCache exploits the highly-charged state of these
recently-accessed rows to lower the latency for
later accesses to them.
NUAT observes that \emph{recently-refreshed} cells are in highly-charged
state, and thus it lowers the latency for accesses to recently-refreshed
rows.
Prior to activating a DRAM row, both ChargeCache and NUAT determine whether
the target row is in a highly-charged state. If so, the memory controller
uses reduced \trcd and \tras timing parameters to perform a low latency
access.



In this section, we evaluate whether or not the expected latency reduction
effect of these two works is observable in existing DRAM modules, using
SoftMC. We first describe our methodology for evaluating the improvement
in the \trcd and \tras timing parameters. We then show the results we
obtain using SoftMC, and discuss our observations. 

\subsubsection{Evaluating DRAM Latency with SoftMC}
\label{sec:trcd_test}
\label{sec:tras_test}

In our experiments, we use 24 DDR3 chips (i.e., three
SO-DIMMs\ch{~\cite{jedec-sodimm}}) from three
major \ch{manufacturers}. To stress DRAM reliability and maximize the amount of cell
charge leakage, we raise the test temperature to 80$^{\circ}$C
(significantly higher than the common-case operating range of
35-55$^{\circ}$C~\cite{lee-hpca2015}) by enclosing our FPGA infrastructure
in a temperature-controlled heat chamber (see
Figure~\ref{fig:softmc_overview}). For all experiments, the temperature
within the heat chamber was maintained within 0.5$^{\circ}$C of the target
80$^{\circ}$C temperature.

To study the impact of charge variation in cells on access latency, which
is dominated by the \trcd and \tras timing
parameters\ch{~\cite{lee-hpca2013, lee-hpca2015, chang2016understanding,
kim-isca2012}}, we perform experiments \ch{on existing DRAM chips} to test
the headroom for reducing these parameters.
In our experiments, we vary one of the two timing parameters, and test
whether the original data can be read back correctly with the reduced
timing. If the data that is read out contains errors, this indicates that
the timing parameter \ch{\emph{cannot}} be reduced to the tested value without inducing
errors in the data. We perform the tests using a variety of data patterns
(e.g., 0x00, 0xFF, 0xAA, 0x55) because 1) different DRAM cells store
information (i.e., 0 or 1) in different states (i.e., charged or
empty)~\cite{liu2013experimental} and 2) we would like to stress DRAM
reliability by increasing the interference between adjacent
bitlines\ch{~\cite{khan2016parbor, liu2013experimental, khan2016case,
patel2017reach, khan2017detecting, khan-sigmetrics2014}}. We also
perform tests using different refresh intervals, to study whether the
variation in charge leakage increases significantly if the time between
refreshes increases.

\begin{figure}[ht]
        \centering

        \subfloat[\trcd Test] {
                \includegraphics[width=0.885\linewidth]{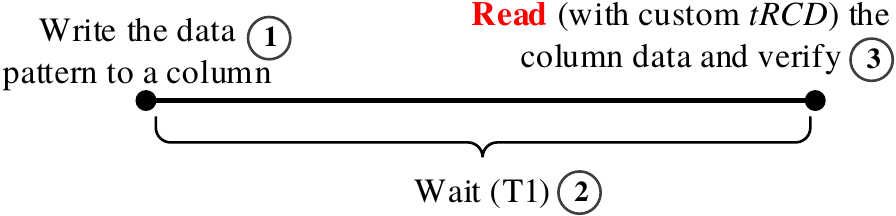}%
                \label{fig:trcd_test}
        }

        \subfloat[\tras Test] {
                \includegraphics[width=0.885\linewidth]{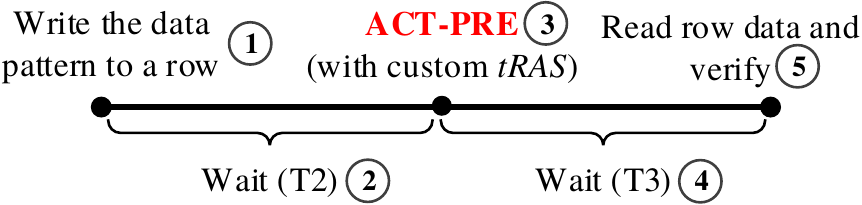}%
                \label{fig:tras_test}
        }
	\caption{Timelines that illustrate the methodology for testing the
improvement of (a) \trcd and (b) \tras on highly-charged DRAM cells.
    Reproduced from~\cite{hassan2017softmc}\ch{.}}
        \label{fig:nuat_tests}
\end{figure}

\textbf{tRCD Test.}
We measure how highly-charged cells affect the \trcd timing
parameter\ch{~(i.e., how long the controller needs to wait after a row
activation \chs{command is sent} to safely perform read and write
operations on the row)}, by using a custom \trcd value to read data from a
row to which we previously wrote a reference data pattern. We adjust the
time between writing a reference data pattern and performing the read, to
vary the amount of charge stored within the cells of a row. In
Figure~\ref{fig:trcd_test}, we show the command sequence that we use to
test whether recently-refreshed DRAM cells can be accessed with a lower
\trcd, compared to cells that are close to the end of the refresh interval.
We perform the write and read operations to each DRAM row one column at a
time, to ensure that each read incurs the \trcd latency. First (\circled{1}
in Figure~\ref{fig:trcd_test}), we perform a reference write to the DRAM
column under test by issuing \cmdact, \cmdwrite, and \cmdprech successively
with the \emph{default} DRAM timing parameters. Next (\circled{2}), we wait
for the duration of a time interval~(\textbf{T1}), which is the refresh
interval in practice, to vary the charge contained in the cells. When we
wait longer, we expect the target cells to have less charge at the end of
the interval. We cover a wide range of wait intervals, evaluating values
between 1 and \SI{512}{\milli\second}. Finally (\circled{3}), we read the
data from the column that we previously wrote to and compare it with the
reference pattern. We perform the read with the custom \trcd value for that
specific test. We evaluate \trcd values ranging from 3 to 6 (default)
cycles. Since a \trcd of 3 cycles produced errors in \ch{\emph{every}} run,
we did not perform any experiments with a lower \trcd.

We process multiple rows in an interleaved manner (i.e., we write to multiple
rows, wait, and then verify their data one after another) in order to further
stress the reliability of DRAM~\cite{lee-hpca2015}. We repeat this process for
all DRAM rows to evaluate the entire memory module.

\begin{figure*}[!hb]
        \centering
        \subfloat[Module A] {
                \includegraphics[height=1.42in]{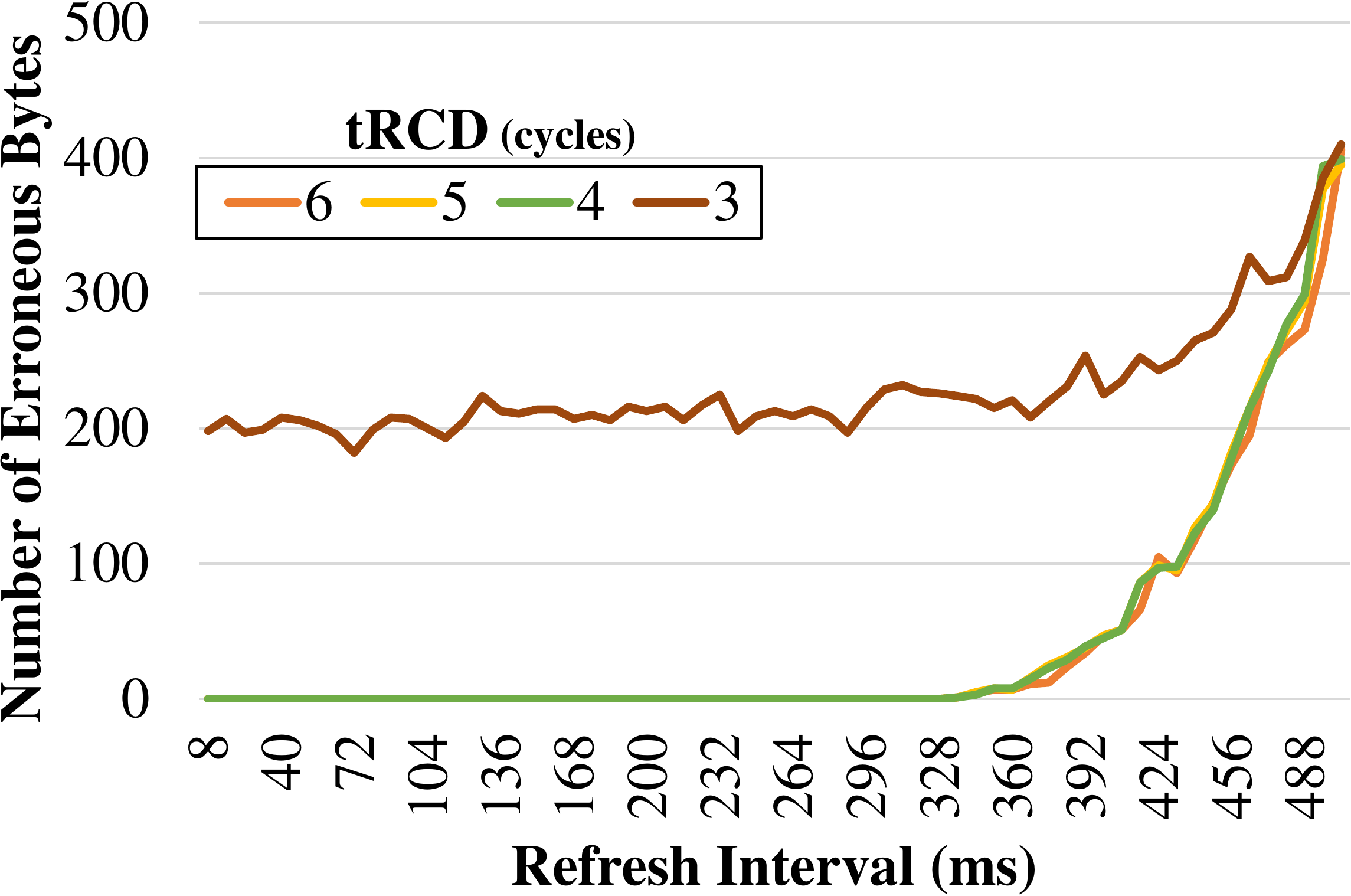}
                \label{fig:trcd_res1}
        }
        \subfloat[Module B] {
                \includegraphics[height=1.42in]{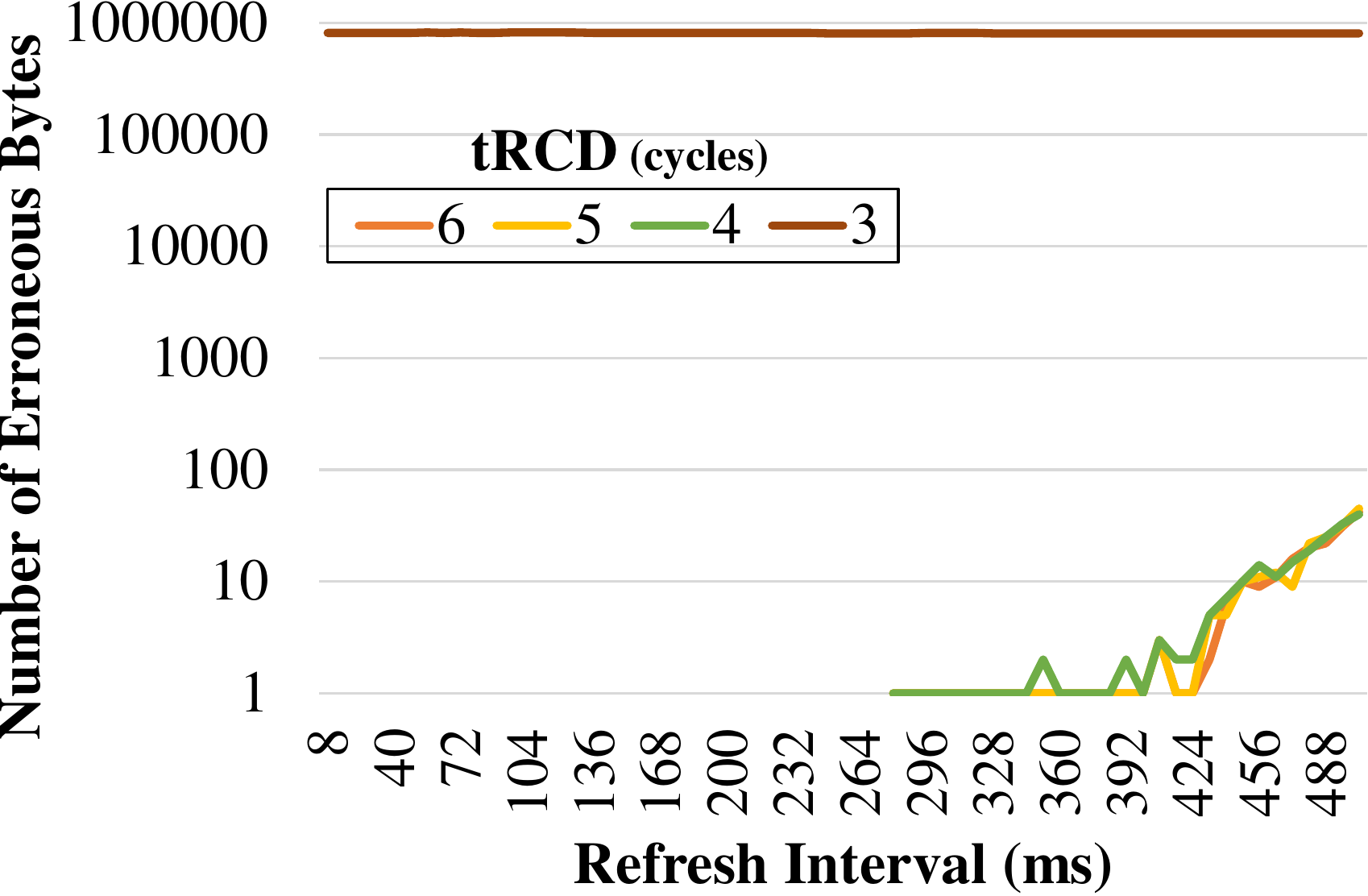}
                \label{fig:trcd_res2}
        }
        \subfloat[Module C] {
                \includegraphics[height=1.42in]{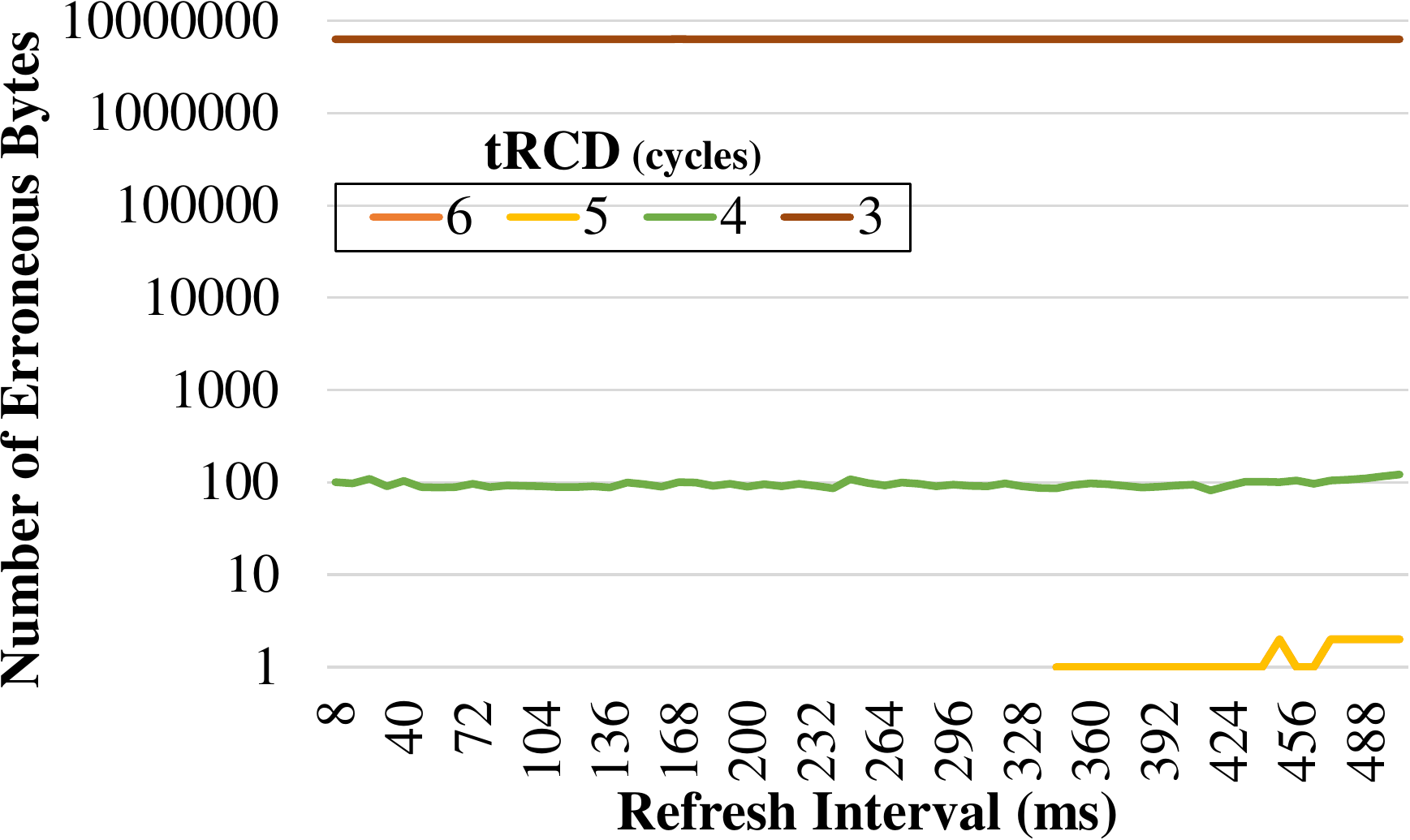}
                \label{fig:trcd_res3}
        }
	\caption{Effect of reducing \trcd on the number of errors at various
    refresh intervals. Reproduced from~\cite{hassan2017softmc}} \label{fig:trcd_test_results}
\end{figure*}

\begin{figure*}[!hb]
        \centering
        \subfloat[Module A] {
                \includegraphics[height=1.4in]{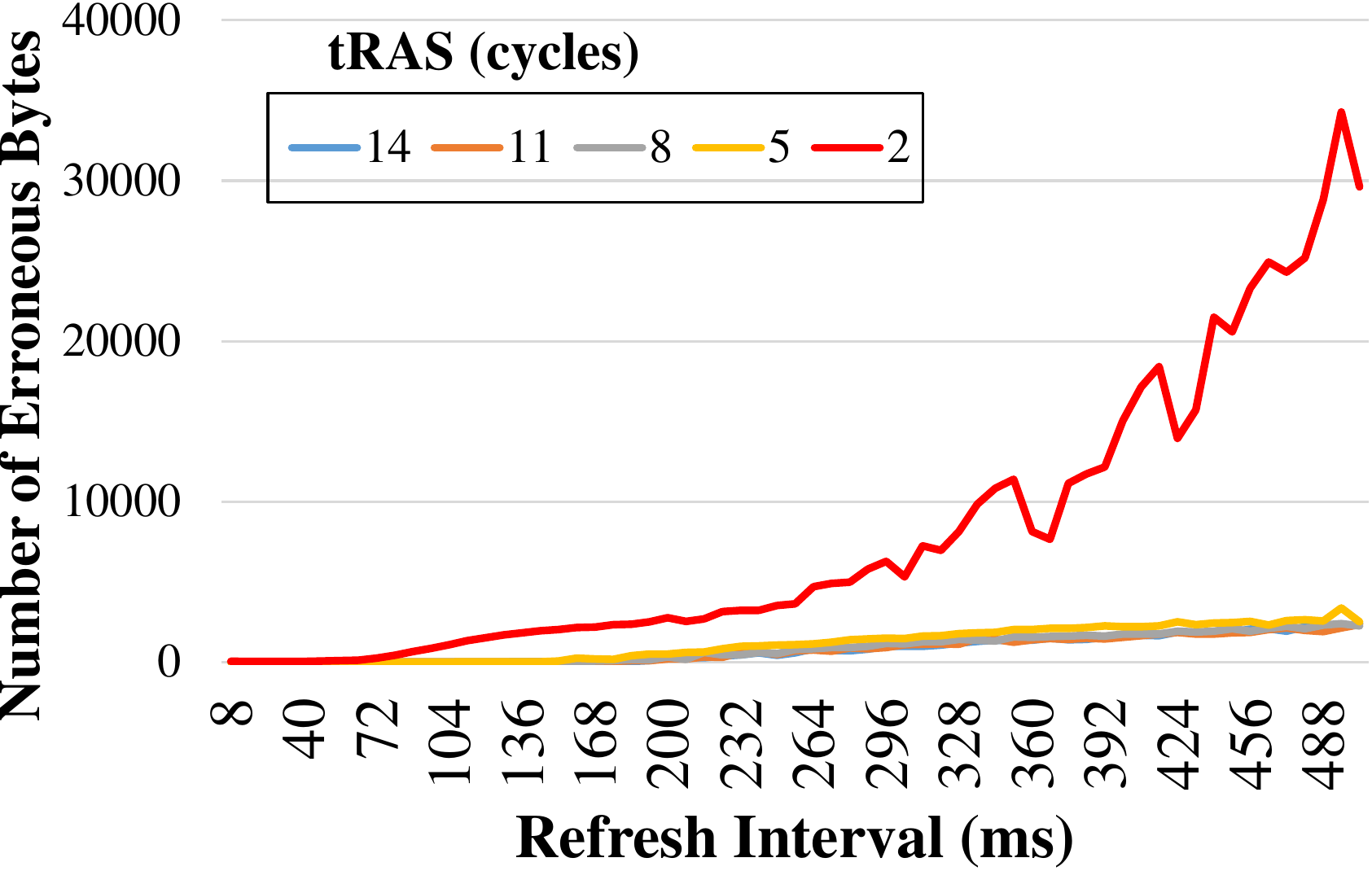}
                \label{fig:tras_res1}
        }
        \subfloat[Module B] {
                \includegraphics[height=1.4in]{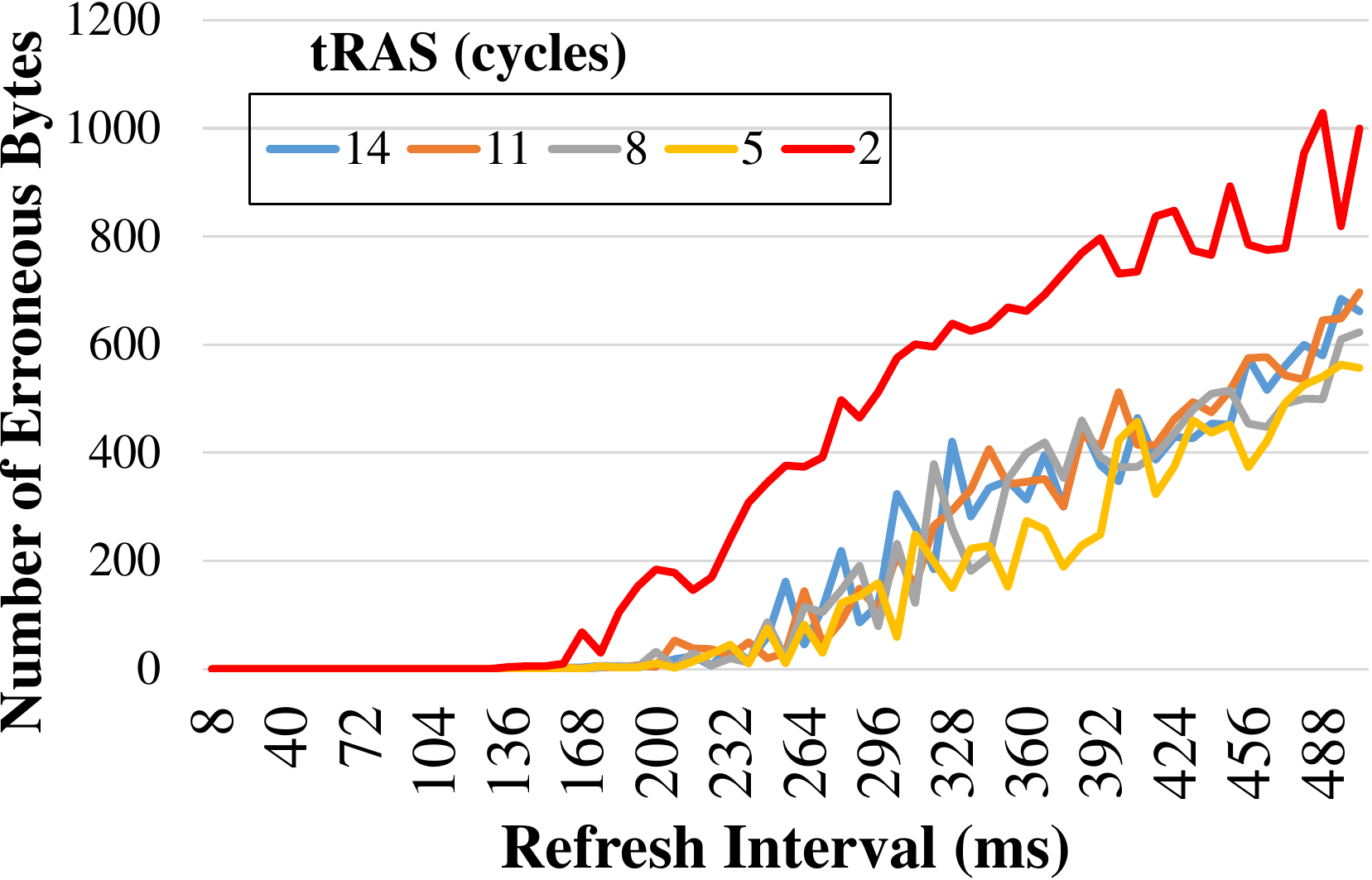}
                \label{fig:tras_res2}
        }
        \subfloat[Module C] {
                \includegraphics[height=1.4in]{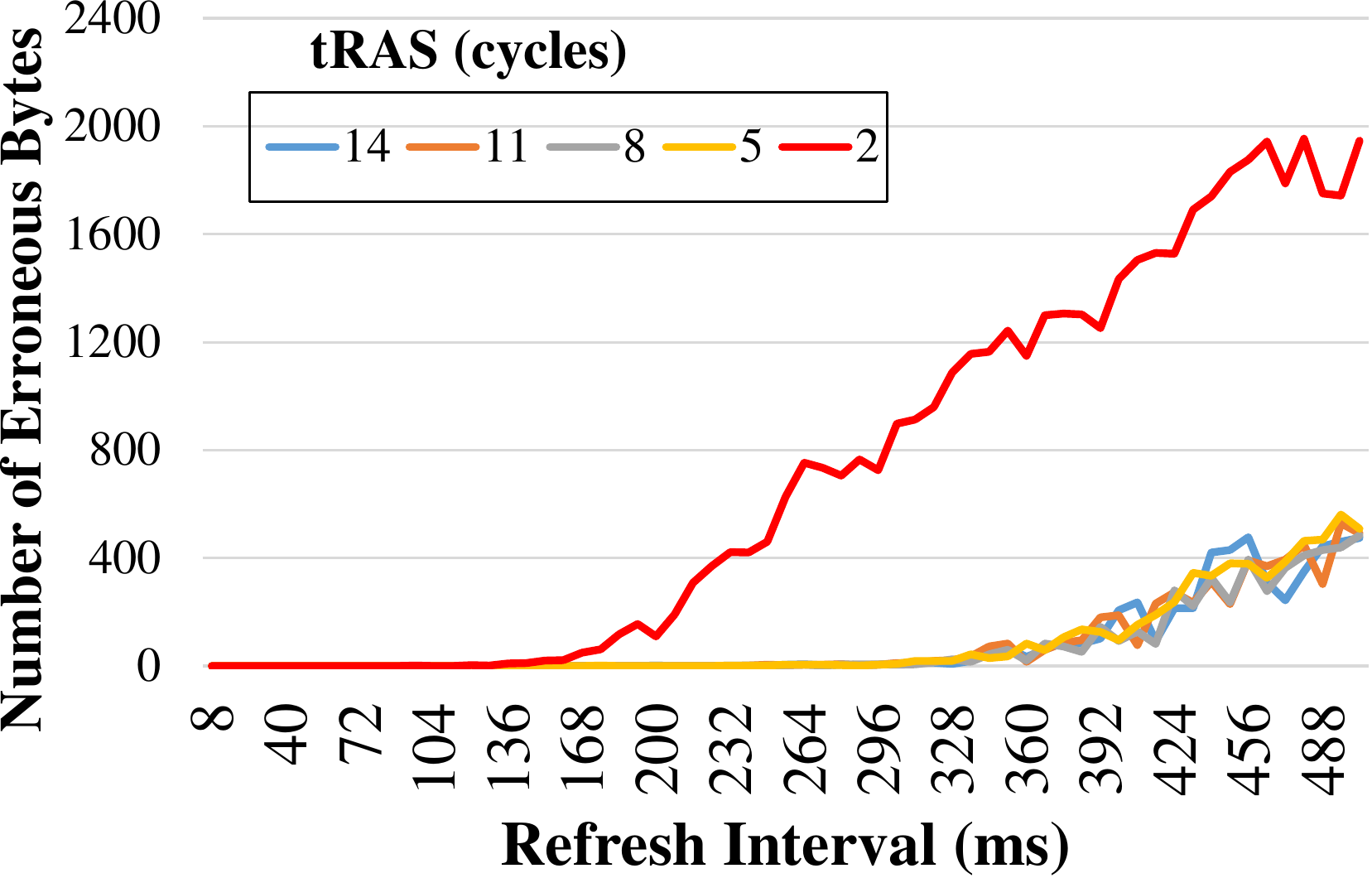}
                \label{fig:tras_res3}
        }
        \caption{Effect of reducing \tras on the number of errors at
        various refresh intervals. Reproduced
        from~\cite{hassan2017softmc}\ch{.}} 
        \label{fig:tras_test_results}
\end{figure*}

\textbf{tRAS Test.} 
We measure the effect of accessing highly-charged rows on the \tras timing
parameter\ch{~(i.e., the time that the controller needs to wait after a row
activation \chs{command is sent} to safely start precharging the row)} by issuing the \cmdact and
\cmdprech commands, with a custom \tras value, to a row. We check if that
row still contains the same data that it held before the \cmdact-\cmdprech
command pair was issued. Figure~\ref{fig:tras_test} illustrates the
methodology for testing the effect of the refresh interval on \tras. First
(\circled{1}), we write the reference data pattern to the selected DRAM row
with the default timing parameters. Different from the \trcd test, we write
to \emph{every column} in the open row (before switching to another row) to
save cycles by eliminating a significant amount of \cmdact and \cmdprech
commands, thereby reducing the testing time. Next (\circled{2}), we wait
for the duration of time interval \textbf{T2}, during which the DRAM cells
lose a certain amount of charge. To refresh the cells (\circled{3}), we
issue an \cmdact-\cmdprech command pair associated with a custom \tras
value. 
When the \cmdact-\cmdprech pair is issued, the charge in the cells of the
target DRAM row may not be fully restored if the wait time is too long or
the \tras value is too short, potentially leading to loss of data. 
Next (\circled{4}), we wait again for a period of time \textbf{T3} to allow
the cells to leak a portion of their charge. Finally (\circled{5}), we read
the row using the default timing parameters and test whether it still
retains the correct data.  Similar to the \trcd test, to stress the
reliability of DRAM, we simultaneously perform the \tras test on multiple
DRAM rows.


We would expect, from this experiment, that the data is likely to maintain
its integrity when evaluating reduced \tras with \emph{shorter} wait times
(\textbf{T2}). 
\chs{This is because when}
\ch{\textbf{T2} is short, a DRAM cell \chI{would} lose only \chs{a \emph{small}} amount of its
charge. Thus, there \chI{would} be more room for reducing \tras, as the cell \chI{would}
already contain \chs{a \emph{higher}} amount of charge prior to the row activation.} 
\chs{The higher amount of charge \chI{would allow} us to safely reduce \tras by a
larger amount.}
In
contrast, we would expect failures to be more likely when using a reduced
\tras with a \ch{\emph{longer}} wait time, because the cells would have a
low amount of charge that is not enough to reliably reduce \tras.

\subsubsection{Results}

We analyze the results of the \trcd and \tras tests, for 24 real
DRAM chips from different vendors, using the test programs
detailed in Section~\ref{sec:trcd_test}. We evaluate \trcd values
ranging from 3 to 6 cycles, and \tras values ranging from 2 to 14
cycles, where the maximum number for each is the default timing
parameter value. For both tests, we evaluate refresh intervals
between 8 and \SI{512}{\milli\second} and measure the number of
observed errors during each experiment.  

Figures~\ref{fig:trcd_test_results}
and~\ref{fig:tras_test_results} depict the results for the \trcd
test and the \tras test, respectively, for three DRAM modules
(each from a different DRAM vendor). We make three major
observations:

\emph{(i)} \emph{Within the duration of the standard refresh interval
(\SI{64}{\milli\second}), DRAM cells do not leak a sufficient amount of
charge to have a negative impact on DRAM access latency.}\footnote{Other
studies have shown methods to take advantage of the fact that latencies can
be reduced without incurring errors~\cite{lee-hpca2015,
chang2016understanding}.} For refresh intervals less than or equal to
\SI{64}{\milli\second}, we observe little to no variation in the number of
errors induced.  Within this refresh interval range, depending on the \trcd
or \tras value, the errors generated are either zero or a constant number.
We make the same observation in both the \trcd and \tras tests for all
three DRAM modules.

For all the modules tested, using different data patterns and
stressing DRAM operation with temperatures significantly higher
than the common-case operating conditions, we can significantly
reduce \trcd and \tras parameters, without observing any errors. We
observe errors only when \trcd and \tras parameters are too small to
correctly perform the DRAM access, regardless of the charge amount of
the accessed cells.



\emph{(ii)} \emph{The large safety margin employed by the
manufacturers protects DRAM against errors even when
accessing DRAM cells with low latency.} We observe no change in
the number of induced errors for \trcd values less than the
default of 6 cycles (down to 4 cycles in modules A and B, and 5
cycles in module C). We observe a similar trend in the \tras
test: \tras can be reduced from the default value of 14
cycles to 5 cycles without increasing the number of induced
errors for \emph{any} refresh interval. 

We conclude that even at temperatures much higher than typical operating
conditions, there exists a large safety margin for access latency in
existing DRAM chips. This demonstrates that DRAM cells are much
\emph{stronger} than their \ch{datasheet} timing specifications indicate.\footnote{Similar
observations were made by prior work~\cite{lee-hpca2015,
chandrasekar2014exploiting, chang2016understanding}.} In other words, the
timing margin in most DRAM cells is very large, given the existing timing
parameters.

\emph{(iii)} \emph{The expected effect of ChargeCache and
NUAT, that highly-charged cells can be accessed with lower
latency, is slightly observable only when very long refresh
intervals are used.} For each of the tests, we observe a
significant increase in the number of errors at refresh intervals
that are much higher than the typical refresh interval of
\SI{64}{\milli\second}, demonstrating the variation in
charge held by each of the DRAM cells. Based on the
assumptions made by ChargeCache and NUAT, we expect that when
lower values of \trcd and \tras are employed, the error rate
should increase more rapidly.  However, we find that for all but
the minimum values of \trcd and \tras (and for \mbox{\trcd= 4} for module
C), the \trcd and \tras latencies have almost no impact on the
error rate.

We believe that the reason we cannot observe the expected
latency reduction effect of ChargeCache and NUAT on existing DRAM
modules is due to the internal behavior of existing DRAM chips
that does not allow latencies to be reduced beyond a certain
point: 
we cannot \emph{externally control} when the sense amplifier gets enabled,
since this is dictated with a fixed latency internally, regardless of the
charge amount in the cell. The sense amplifiers are enabled only after
charge sharing, which starts by enabling the wordline and lasts until
sufficient amount of charge flows from the activated cell into the
bitline\ch{~\cite{lee-hpca2013, kim-isca2012, seshadri2013, chang2016,
seshadri2017ambit, seshadri2015fast}},
is expected to complete. Within existing DRAM chips, the expected charge
sharing latency (i.e., the time when the sense amplifiers get enabled) is
\ch{\emph{not}} represented by a timing parameter managed by the memory controller.
Instead, the latency is controlled internally within the DRAM using a fixed
value~\cite{tomishima2016dram, keeth2008dram}.
ChargeCache and NUAT require that charge sharing completes in less time,
and the sense amplifiers get enabled faster for a highly-charged cell.
However, since existing \chI{DRAM chips} provide no way to control the time it takes
to enable the sense amplifiers, we cannot harness the potential latency
reduction possible for highly-charged cells~\cite{tomishima2016dram}.
Reducing \trcd affects the time spent \emph{only after charge sharing}, at
which point the bitline voltages exhibit similar behavior regardless of the
amount of charge initially stored within the cell.  Consequently, we are
unable to observe the expected latency reduction effect of ChargeCache and
NUAT by simply reducing \trcd, even though we believe that the mechanisms
are sound and can reduce latency (assuming the behavior of DRAM chips is
modified). If the DDR interface exposes a method of controlling the time it
takes to enable the sense amplifiers in the future, SoftMC can be easily
modified to use the method and fully evaluate the latency reduction effect
of ChargeCache and NUAT.

\textbf{Summary.}
Overall, we make two major conclusions from the implementation and
experimental results of our DRAM latency experiments. First, SoftMC provides a
simple and easy-to-use interface to quickly implement tests that
characterize modern DRAM chips. Second, SoftMC is an effective
tool to validate or refute the expected effect of existing or new
mechanisms on existing DRAM chips. 


%% file: sections/related_work.tex

\section{Related Work}
\label{sec:relatedk}

No prior DRAM testing infrastructure provides both
\emph{flexibility} and \emph{ease of use} properties, which are critical
for enabling widespread adoption of the infrastructure. Three
different kinds of tools/infrastructure are available today for
characterizing \chI{the behavior of real DRAM chips}. As we will describe,
each kind of tool has some shortcomings. SoftMC \ch{eliminates}
\emph{all} of these shortcomings \ch{and provides the first open-source
DRAM testing infrastructure that is publicly available~\cite{safarigithub}}.

\textbf{Commercial Testing Infrastructures.}
A large number of commercial DRAM testing platforms (e.g.,~\cite{advantest,
nickel, teradyne, futureplus}) are available in the market. Such platforms
are optimized for \ch{test} throughput (i.e., to test as many DRAM chips as
possible in a given time period), and generally apply a \emph{fixed test
pattern} to the units under test. Thus, since they lack support for
flexibility in defining the test routine, these infrastructures are not
suitable for detailed DRAM characterization where the goal is to
investigate new issues and new ideas. Furthermore, such testing
equipment is usually quite expensive, which makes these
infrastructures an impractical option for research in academia.
Industry may also have internal DRAM development and testing tools,
but, to our knowledge, these are proprietary and are unlikely to be made
openly available.

We \ch{design} SoftMC to be a low-cost (i.e., free) and flexible
open-source alternative to commercial testing equipment that can enable new
research directions and mechanisms. For example, prior
work~\cite{yang2015random} recently proposed a random command pattern
generator to validate DRAM chips against uncommon yet supported (according
to JEDEC specifications) DDR command patterns. Using the test patterns on
commercial test equipment, this work demonstrates that specific sequences
of commands introduce failures in current DRAM chips\ch{~(\chs{e.g.,
an \cmdact followed by a \cmdprech, without any \cmdread or \cmdwrite commands in between,
results in \emph{future} accesses reading incorrect data in} some DRAM
devices)}. SoftMC flexibly supports the ability to issue an arbitrary
command sequence, and therefore can be used as a low-cost method for
validating DRAM chips against problems that arise due to command ordering.

\textbf{FPGA-Based Testing Infrastructures.}
Several prior works \chI{propose} FPGA-based DRAM testing
infrastructures~\cite{huang2000fpga, hou2013fpga, keezer2015fpga}.
Unfortunately, all of them lack flexibility and/or a simple user
interface, and none are open-source. The
FPGA-based infrastructure proposed by Huang et al.~\cite{huang2000fpga}
provides a high-level interface for developing DRAM tests, but the
interface is limited to defining only data patterns and march algorithms
for the tests. Hou et al.~\cite{hou2013fpga} propose an FPGA-based test
platform whose capability is limited to analyzing only the data retention
time of the DRAM cells. Another work~\cite{keezer2015fpga}
develops a custom memory testing board with an FPGA chip,
specifically designed to test memories at a very high data rate. However,
it requires low-level knowledge to develop FPGA programs, and even then
offers only limited flexibility in defining a test routine. On the other
hand, SoftMC \ch{provides} \emph{full control over all DRAM commands}
using a high-level \emph{software interface}, and it is
open-source.

PARDIS~\cite{bojnordi2012pardis} is a reconfigurable logic
(e.g., FPGA) based programmable memory controller meant to be implemented inside
microprocessor chips. PARDIS is capable of optimizing memory scheduling
algorithms, refresh operations, etc. at run-time based on application
characteristics, and can improve system performance and efficiency. However, it does not
provide programmability for DRAM commands and timing parameters, and therefore
cannot be used for detailed DRAM characterization.


\textbf{Built-In Self Test (BIST).}
A BIST mechanism (e.g,\ch{~\cite{you1985self, querbach2014reusable,
querbacharchitecture, bernardi2010programmable, yang2015hybrid,
itoh2013vlsi}}) is
implemented inside the DRAM chip to enable fixed test patterns and
algorithms. Using such an approach, DRAM tests can be performed
faster than with other testing platforms. However, BIST has two
major flexibility issues, since the testing logic is hard-coded into the
hardware: \emph{(i)} BIST offers only a limited number of tests that are
fixed at hardware design time. \emph{(ii)} A limited set of DRAM
chips, which come with BIST support, can be tested. In contrast,
SoftMC allows for the implementation of a wide range of DRAM test
routines and supports any off-the-shelf DRAM chip that is compatible with the
DDR interface.



\textbf{Other Related Work\ch{.}} Although no prior work provides an open-source
DRAM testing infrastructure similar to SoftMC, infrastructures for testing
other types of memories have been developed. Cai et al.\chI{~\cite{cai2011fpga, cai2017error,
cai.bookchapter.arxiv17, cai.procieee.arxiv17}}
\chI{develop} a platform for characterizing NAND flash memory. They propose a
flash controller, implemented on an FPGA, to quickly characterize error
patterns of existing flash memory chips. They expose the functions of
the flash translation layer (i.e., the flash chip interface) to
the software developer via the host machine connected to the FPGA board,
similar to how we expose the DDR interface to the user in SoftMC. Many
works\chs{~\cite{luo2016enabling, cai2015read, luo2015warm, cai2015data,
cai2014neighbor, cai2013program, cai2013error, cai2013threshold,
cai2012flash, cai2012error, cai2017vulnerabilities, cai2017error,
fukami2017improving, cai.bookchapter.arxiv17, cai.procieee.arxiv17,
luo.hpca18}} use this flash memory
testing infrastructure to study various aspects of flash chips.

Our prior works\ch{~\cite{chang2016understanding, khan2016parbor, khan2016case,
kim2014flipping, khan-sigmetrics2014, liu2013experimental,
lee-hpca2015, lee2017design}} \chI{develop and use} FPGA-based infrastructures for a wide
range of DRAM studies. Liu et al.~\cite{liu2013experimental} and Khan et
al.~\cite{khan-sigmetrics2014} \chI{analyze} the data retention behavior of
modern DRAM chips and proposed mechanisms for mitigating retention
failures. Khan et al.~\cite{khan2016parbor, khan2016case} \chI{study}
data-dependent failures in DRAM, and developed techniques for efficiently
detecting and handling them. Lee et al.~\cite{lee-hpca2015,
lee2017design} \chI{analyze} latency characteristics of modern DRAM chips and
\chI{propose} mechanisms for latency reduction. Kim et
al.~\cite{kim2014flipping} \chI{discover} a new reliability issue in existing
DRAM, called \emph{RowHammer}, which can lead to security
breaches\ch{~\cite{seaborn2015exploiting, seaborn2015exploiting2,
rowhammerjava2015, van2016drammer, xiao2016one, razavi2016flip,
mutlu2017rowhammer}}. Chang et
al.~\cite{chang2016understanding} \chI{use} SoftMC to characterize latency
variation across DRAM cells for fundamental DRAM operations (e.g.,
activation, precharge).  SoftMC evolved out of these previous
infrastructures, to address the need to make the infrastructure flexible
and easy to use.

\ch{
    Recently, Chang et al\chs{.}~\cite{chang2017understanding} \chI{extend} SoftMC
    with the capability to change the array voltage of DRAM chips\chs{,
    such that SoftMC can be used to \chI{evaluate} the} trade-offs \chs{between
    voltage, latency, and reliability \chI{in modern DRAM chips}}. 

    Sukhwani et al.\ \chI{propose} ConTutto~\cite{sukhwani2017contutto}, which is a
    recent work that builds an FPGA-based platform for evaluating different
    memory technologies and new mechanisms on existing server systems.
    ConTutto is an extender board, which plugs into the DDR3 module slot of
    a server machine. On the board, an FPGA chip manages the communication
    between the server machine and the memory, which is connected to the
    other end of the ConTutto board. Using ConTutto, any type of memory
    that can be attached to the ConTutto board can potentially be used in
    existing systems\chs{, as part of main memory,} by using the FPGA as a translator between the two
    interfaces, i.e., \chs{between the} DDR3 interface to the server and the interface
    of the memory attached to the ConTutto board. Although
    ConTutto can be used as a prototyping platform to evaluate different
    memory technologies and mechanisms on existing systems, it is \chI{\emph{not}
    practical or flexible enough} to use for testing memories \chs{for two reasons}. 
    First, the operating system needs to ensure that it does not allocate
    \chs{application} data to the memory that is being tested, as the data
    could be destroyed during a testing procedure. Second, the memory that
    is connected to ConTutto is accessed using load/store instructions,
    which does \chI{\emph{not}} provide the flexibility of testing \chs{the memory} at
    \chs{the} memory command level. In contrast, \chI{(1)~the} memory in SoftMC is
    not a part of the main memory of the host machine\chs{,} and \chI{(2)~SoftMC}
    provides a high-level software interface for directly issuing commands
    to the memory.  \chI{These design choices enable many tests that are not otherwise} possible or
    practical to implement using load/store instructions.}

We conclude that prior work lacks either the flexibility or
the ease-of-use properties that are critical for performing
detailed DRAM characterization. To fill the gap left by current
infrastructures, we introduce an open-source DRAM testing
infrastructure, SoftMC, that fulfills these two properties.

%% file: sections/significance.tex
\section{Significance}


Computing systems typically use DRAM-based memories as main memory since
DRAM provides large capacity and high performance. As the process
technology scales down, DRAM technology faces challenges that impact its
reliability and performance\ch{~\cite{mutlu2013memory,
mutlu2017rowhammer}}. Our HPCA \ch{2017} paper~\cite{hassan2017softmc} introduces
SoftMC, a new DRAM characterization infrastructure that is flexible and
practical to use. We release SoftMC as a publicly-available open-source
tool~\cite{safarigithub}. In this section, we discuss the significance of
our work by \ch{describing} its novelty and long-term impact. \ch{We also
discuss various future research directions in which SoftMC can be extended
and \chs{applied}.}

\subsection{Novelty}


\chI{As we describe in Section~\ref{sec:relatedk}, no}
prior DRAM testing infrastructure provides both \emph{flexibility} and
\emph{ease of use} properties, which are critical for enabling widespread
adoption of the infrastructure. Three different kinds of
tools/infrastructures are available today for characterizing DRAM behavior,
where each kind of tool has some shortcomings. \chs{We discuss these tools
and their shortcomings in Section~\ref{sec:relatedk}.}
In contrast \ch{to all these works}, SoftMC allows for the
implementation of a wide range of DRAM test routines and supports any
off-the-shelf DRAM chip that is compatible with the DDR interface.
\ch{SoftMC is also \chI{the first DRAM characterization tool that is} 
freely available to public~\cite{ramulatorgithub}.}

%

\subsection{\ch{Research Directions Enabled by SoftMC}}

\input{sections/future_directions}

%% file: sections/future_directions.tex

We believe SoftMC can enable many new studies of the behavior of DRAM and
other memories. We briefly describe several examples in this section.

\sloppypar{ 
\textbf{\chI{Enabling New Studies of DRAM Scaling and Failures}.} The SoftMC DRAM testing
infrastructure can  test any DRAM mechanism consisting of low-level DDR
commands. Therefore, it enables a wide range of characterization and
analysis studies of real DRAM modules that would otherwise not have been
possible without such an infrastructure. We discuss three such example
research directions.  
}

First, as DRAM scales down to smaller technology nodes, it faces key
challenges in both reliability and latency\ch{~\cite{mutlu2013memory,
mandelman2002challenges, kim2005technology, mueller2005challenges,
liu2012raidr, kang-memcon2014, liu2013experimental, khan2016case,
khan2016parbor, chang2016understanding, chang2017understanding,
khan2017detecting, mutlu2017rowhammer}}. Unfortunately, there is no
comprehensive experimental study that characterizes and analyzes the trends
in DRAM cell operations and behavior with technology scaling across various
DRAM generations.  The SoftMC infrastructure can help us answer various
questions to this end: How are the cell characteristics, reliability, and
latency changing with different generations of technology nodes? Do all
DRAM operations and cells get affected by scaling at the same rate?  Which
DRAM operations are getting worse?  

Second, aging-related failures in DRAM can potentially affect the
reliability and availability of systems in the field\ch{~\cite{meza-dsn2015,
schroeder2009dram, mutluresearch, mutlu2013memory}}. However, the causes,
characteristics, and impact of \emph{aging} \chI{in real DRAM devices} have remained largely
unstudied. Using SoftMC, it is possible to devise controlled experiments to
analyze and characterize \chI{DRAM} aging. The SoftMC infrastructure can help us
answer questions such as: How prevalent are aging-related failures? What
types of usage accelerate aging? How can we design architectural techniques
that can slow down the aging process?

Third, prior works show that the failure rate of DRAM modules in large data
centers is significant, largely affecting the cost and downtime in data
centers~\cite{schroeder2009dram, sridharan2013feng, meza-dsn2015,
luo2014characterizing}. Unfortunately, there is no study that analyzes
DRAM modules that have failed in the field to determine the common causes
of failure. Our SoftMC infrastructure can test faulty DRAM modules and help
answer various research questions: What are the dominant types of DRAM
failures at runtime? Are failures correlated to any location or specific
structure in DRAM?  Do all chips from the same generation exhibit the same
failure characteristics?  \chI{Do failures repeat?}

\sloppypar{
\textbf{Characterization of Non-Volatile Memory.}
The SoftMC infrastructure can test any chip compatible 
with the DDR interface. Such a design makes the scope of the chips 
that can be tested by SoftMC go well beyond just DRAM. 
With the emergence of \chI{byte-addressable}
non-volatile memories (e.g., 
\chs{phase-change memory\chI{~\cite{Raoux:2008:PRA,lee-main-memory, lee2009architecting, lee2010phase, meza.weed13, wong.procieee10, ren.micro15, qureshi.isca09, zhou.isca09}}, 
STT-MRAM~\cite{4443191,sttmram-ispass13, naeimi.itj13, meza.weed13}, 
RRAM/memristors~\cite{5607274, wong2012metal, chua.tct71, strukov.nature08}}), 
several vendors are working towards manufacturing 
DDR-compatible non-volatile memory chips at a large scale~\cite{xpoint,everspin}.
When these chips become commercially available, it will be 
critical to characterize and analyze them in order 
to understand, exploit, and/or correct their behavior. We believe that 
SoftMC can be seamlessly used to characterize these chips, and can help 
enable future mechanisms for NVM.
}

SoftMC will hopefully enable other works that build on it in various ways.
For example, future work can extend the infrastructure to enable
researchers to analyze memory scheduling (\chI{e.g., \cite{mutlu2008parallelism,
mutlu2007stall, subramanian2014blacklisting, kim2010atlas, kim2010thread,
muralidhara2011reducing, ebrahimi2011parallel, moscibroda.usenixsecurity07,
frfcfs-patent, fr-fcfs, ipek-isca08, morse-hpca12, ghose2013,
pa-micro08, cjlee-micro09, mutlu-podc08}}) and memory power
management~\cite{david2011memory, deng2011memscale} mechanisms, and allow
them to develop new mechanisms using a programmable memory controller and
real workloads. \ch{SoftMC can also be used as a substrate for developing
in-memory computation platforms and \chs{evaluating} mechanisms \chs{for} in-memory
computation \chI{(e.g., \cite{ahn2015scalable, ahn2015pim,
    hsieh2016accelerating, hsieh2016transparent, boroumand2017lazypim,
    seshadri2017ambit, seshadri2013, seshadri2015gather,
    fraguela2003programming, elliott1999computational,
    draper2002architecture, kang2012flexram, kogge1994execube,
    oskin1998active, patterson1997case, shaw1981non, stone1970logic, 
    kim.bmc18, boroumand.asplos18}).}}

 We conclude that characterization with SoftMC enables a
wide range of research directions in DDR-compatible memory chips (DRAM or
NVM), leading to better understanding of these technologies and helping to
develop mechanisms that improve the reliability and performance of future
memory systems.

%% file: sections/conclusion.tex
\section{Conclusion}
\label{sec:conclusion}

This work introduces the first publicly-available FPGA-based DRAM testing
infrastructure, {\em SoftMC} (Soft Memory Controller), which provides a
programmable memory controller with a flexible and easy-to-use software
interface. SoftMC enables the flexibility to test any standard DRAM
operation and any (existing or new) mechanism comprising of such
operations. It provides an intuitive high-level software interface for the
user to invoke low-level DRAM operations, in order to minimize programming
effort and time. We provide a prototype implementation of SoftMC, and we
have released it publicly as a freely-available open-source
tool~\cite{safarigithub}.

We demonstrate the capability, flexibility, and programming ease of SoftMC
by implementing two example use cases.  Our experimental analyses
demonstrate the effectiveness of SoftMC as a new tool to {\em (i)} perform
detailed characterization of various DRAM parameters (e.g., refresh
interval and access latency) as well as the relationships between them, and
{\em (ii)} test the expected effects of existing or new mechanisms (e.g.,
whether or not highly-charged cells can be accessed faster in existing DRAM
chips).  We believe and hope that SoftMC, with its flexibility and ease of
use, can enable many other studies, ideas and methodologies in the design
of future memory systems, by making memory control and characterization
easily accessible to a wide range of software and hardware developers.



%% file: sections/acknowledgements.tex
\section*{Acknowledgments}
\label{section:acknowledgements}
We thank the reviewers, the SAFARI group members, and Shigeki
Tomishima from Intel for their feedback. We acknowledge the generous
support of Google, Intel, NVIDIA, Samsung, and VMware. This work is
supported in part by NSF grants 1212962, 1320531, and 1409723, the Intel Science
and Technology Center for Cloud Computing, and the Semiconductor Research
Corporation. 